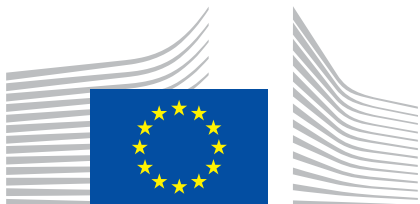
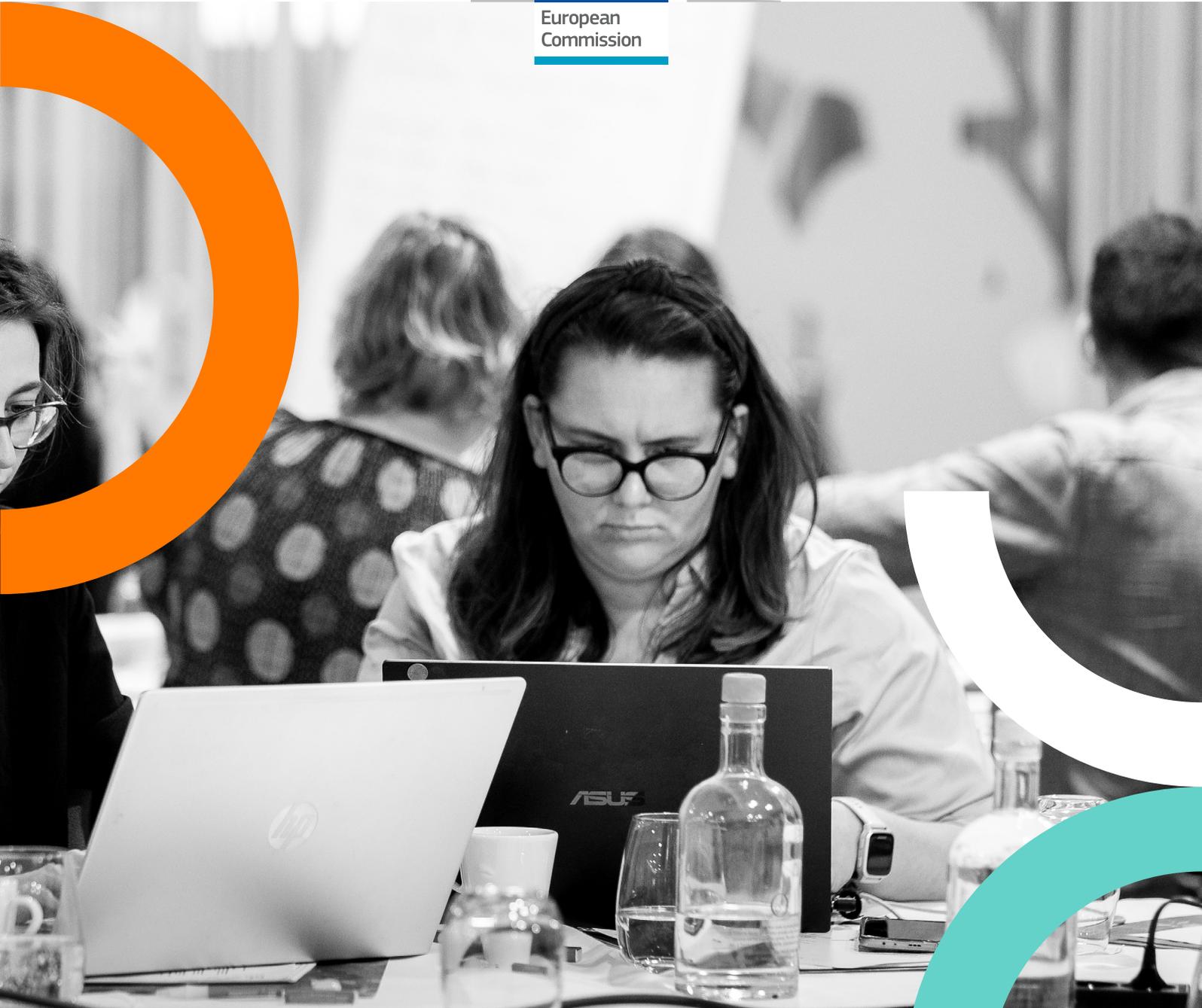

# Use Scenarios & Practical Examples of AI Use in Education

**Briefing report No. 3
by the European Digital Education Hub's squad on artificial intelligence in education**

EUROPEAN DIGITAL EDUCATION HUB

# Content

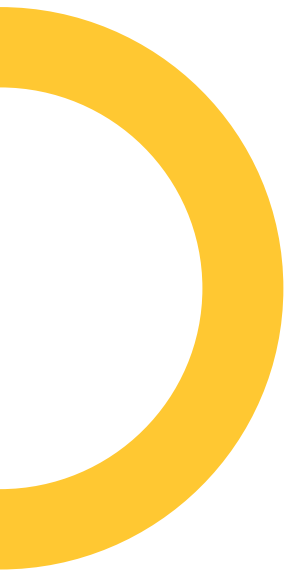
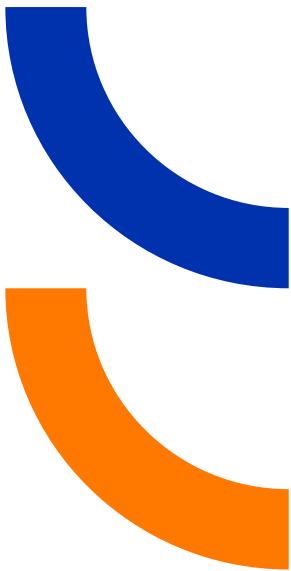
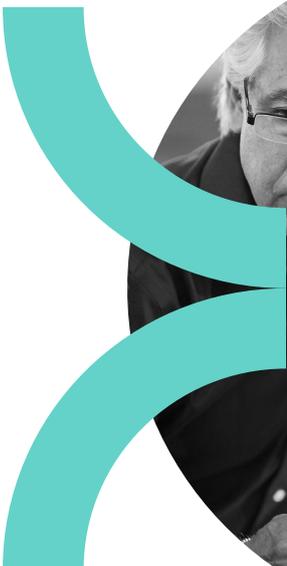
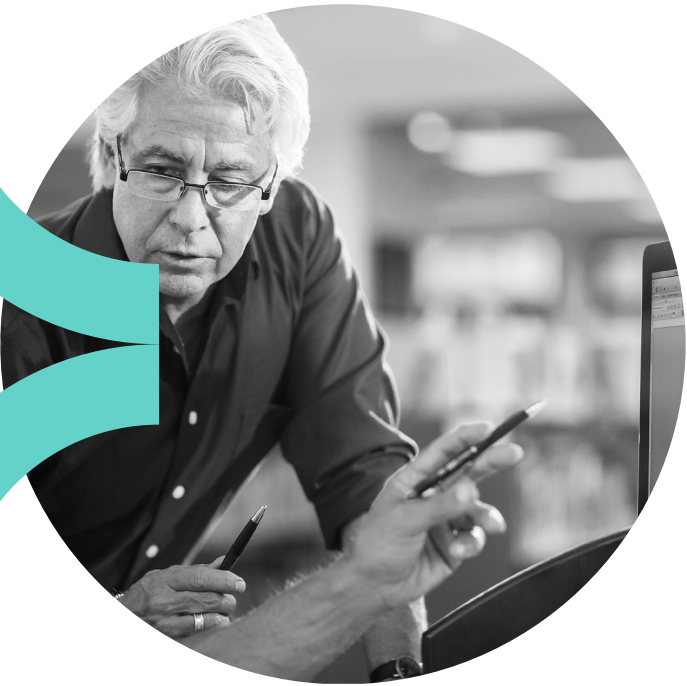

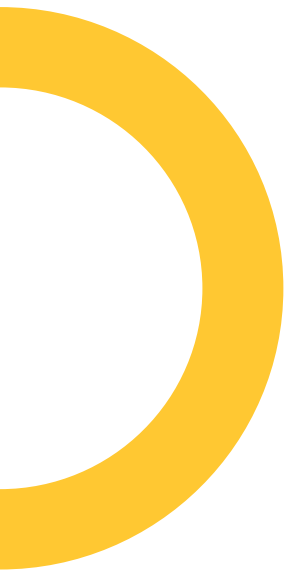
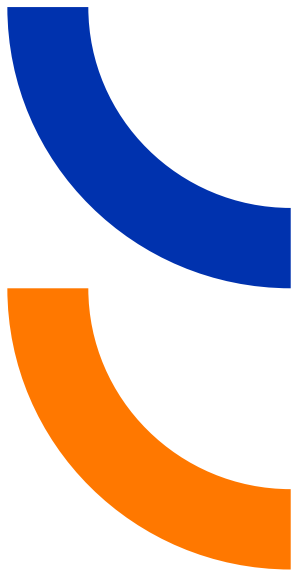
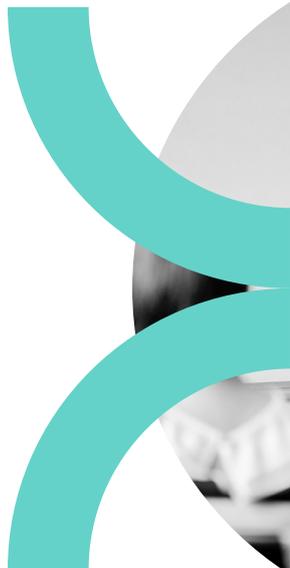
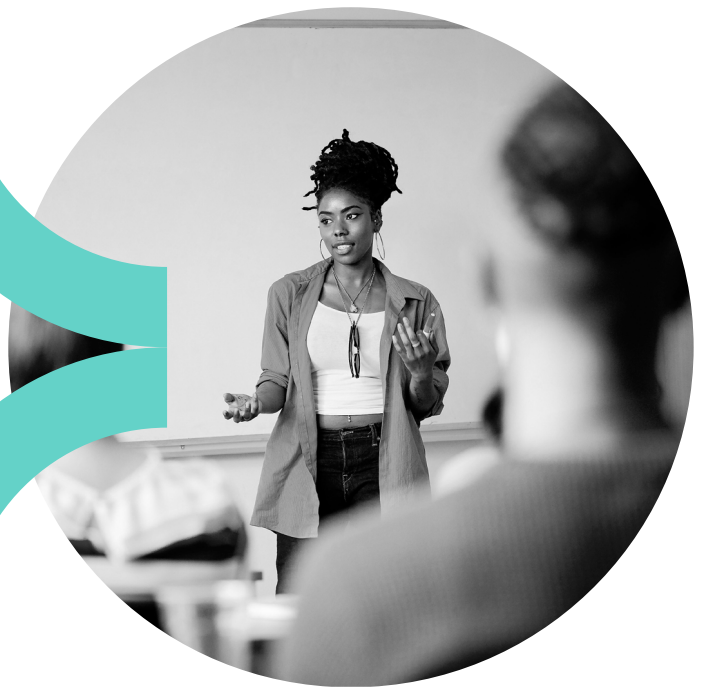

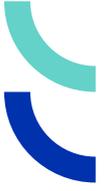

# Introduction and Report Organisation

This report presents a set of use scenarios based on existing resources that teachers can use as inspiration to create their own, with the aim of introducing artificial intelligence (AI) at different **pre-university levels,** and with different goals. The Artificial Intelligence Education field (AIEd) is very active, with new resources and tools arising continuously. Those included in this document have already been tested with students and selected by experts in the field, but they must be taken just as practical examples to guide and inspire teachers' creativity.

The use scenarios have been organised in three main categories, according to the three main approaches followed in AIEd:

Teaching *for* AI entails competences for all citizens, including teachers and learners, to engage confidently, critically, and safely with AI systems to provide them with the necessary knowledge, skills and attitudes to live in a world surrounded and shaped by AI.

Teaching *about* AI is the more technical part, focused on training students in the fundamentals of AI. It is usually part of AI literacy which should comprise both the technological and the human dimensions of AI organised according to the student's age. Knowledge about AI basics is key for preparing students for the labour market, independently of their future careers.

Teaching *with* AI focuses on the application of AI-based tools for educational goals. These types of tools provide autonomous support to students in different aspects of learning, facilitating teachers' work. Moreover, they can also support teachers and institutions in management and supervisory tasks. The main goal in this category is not to understand the technology behind these tools , but to take advantage of the potential of AI to enhance teaching and learning.

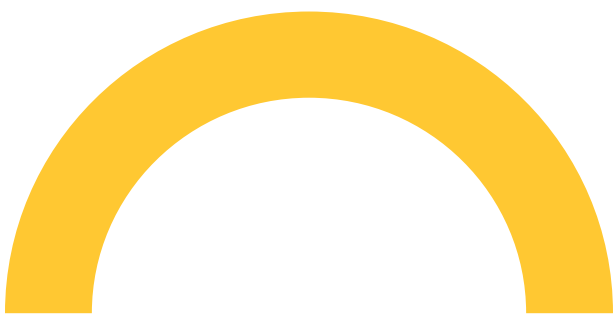
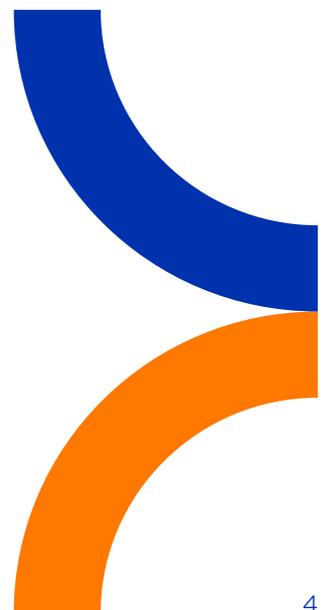

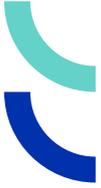

While the third category is quite clear, the difference between the first and second could be confusing for teachers. Teaching *for* AI implies training students in all the AI topics from an AI **user perspective**, rather than an AI **developer perspective**, which will be covered in the Teaching about AI section. We could differentiate these two perspectives with the following key ideas:

- In terms of **curriculum**, teaching *about* AI should be included as specific subjects or courses (or part of them) with a detailed program covering the main AI topics (perception, actuation, reasoning, representation, learning, impact, etc.). **The learning outcomes are more technical and specific,** so before learning about AI, students should receive background training in maths, programming, and other technical knowledge required to properly understand the AI topics from a developer perspective. Teaching *for* AI could be organised in a more transversal manner through embedding it in different courses and areas (e.g., language, history, natural sciences, mathematics, arts). The learning material could be organised as small activities within different subjects (not only technical), or as specific subjects where the AI topics are delivered without relying on deep technical aspects (like programming). Learning for AI does not require a specific background in maths or programming.
- In terms of **methodology**, in teaching *about* AI, students develop simple AI-based solutions by programming them, while in teaching *for* AI, they can focus on analysing existing AI-based applications or tools by using them, understanding the way they work and their impact.
- In terms of **specialisation**, teaching *for* AI is necessary for all students, independently of their area (humanities, science, engineering, arts). Teaching *about* AI could be targeted to technical paths, thinking about those students interested in working as "AI engineers". Hence, **teaching *for* AI is a pre-requisite for educators and learners before moving to teaching about AI.**

The following 3 sections contain selected use scenarios in these categories that exemplify their differences and opportunities at classes.

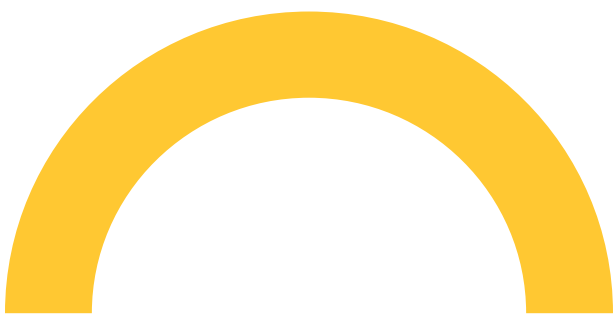
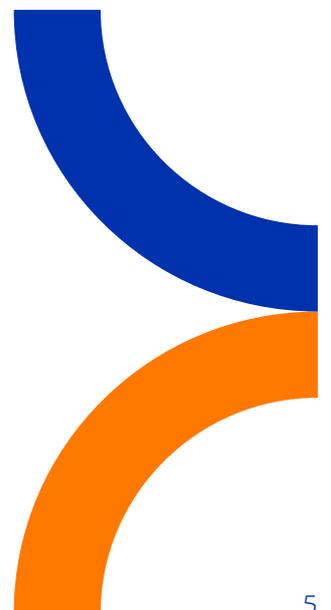

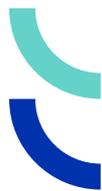

# Teaching *for* AI

In this section we describe projects and learning scenarios that provide activities for teachers to engage students in activities that improve knowledge, skills, and attitudes towards how AI systems are used in today's society, and focus on everyday applications that are driven by AI. To teach for AI use in everyday life, the focus is two-fold:

- to introduce some basic principles to keep in mind when interacting with **common AI applications,** such as virtual assistants and recommendation systems, to mitigate risks related to safety, personal data, privacy and well-being;
- to become aware of **how AI is used in various parts of our society**, e.g., autonomous vehicles, medicine, industry, agriculture, and creating realistic expectations about what AI systems do and do not do.

This is a key topic, and students should be aware of the limits of AI from a formal perspective, mainly understanding the differences with human intelligence.

When creating lesson plans to teach *for* AI, areas that are part of the DigComp framework can become helpful. Especially **understanding AI's impact** on information, data and media literacy is crucial due to disinformation on social media platforms and other new forms of automated AI -generated content that exists on the internet.

AI literacy and digital citizenship are essential topics to cover and should include examples for responsible use of AI and data-driven technologies, with a critical mindset to be aware about the potential biases and limitations of such systems. Here, an important goal is to help people navigate ethical questions related to digital practices – like the question of human autonomy which underpins many of the EU values.

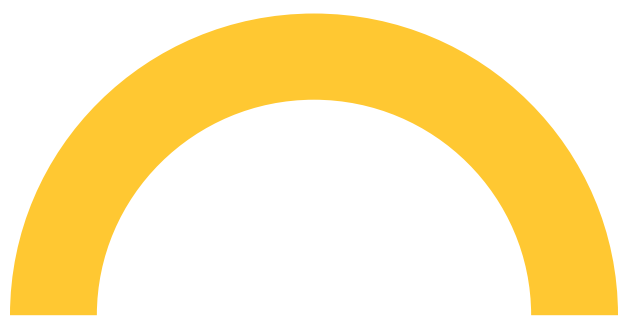
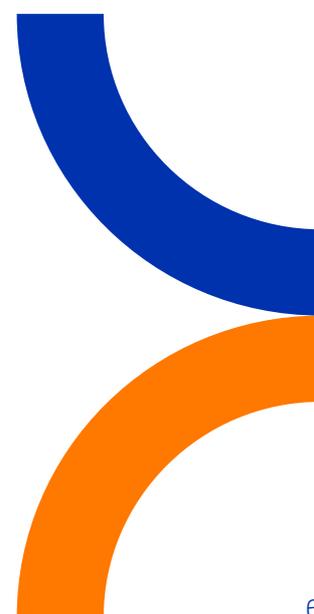

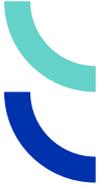

At a certain student age, the inclusion of **legal topics** such as personal data protection (GDPR) and privacy, ethical considerations in data collection, storage, and usage, and bias and fairness in AI algorithms will be important.

It could also be useful to include examples about **AI applications in tools and services,** including using AI-powered tools for productivity, communication, and entertainment, integrating AI services into custom applications using application programming interfaces (APIs), and evaluating AI services for data privacy and security concerns. Similarly, introducing students to methods of exploratory data analysis through descriptive statistics and data distributions, data visualisation techniques and tools such as bar charts, pie charts, and scatter plots, and making data-driven decisions based on analysis and visualisations could be useful as well.

Teaching *for* AI can also include aspects such as understanding the nature of different types of data (structured, unstructured, and semi-structured), data formats (text, images, audio, and video), and data sources (public datasets, APIs, and web scraping), and more technical concepts of AI such as machine learning and deep learning.

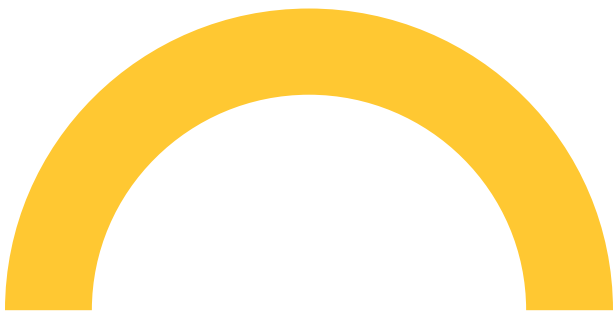
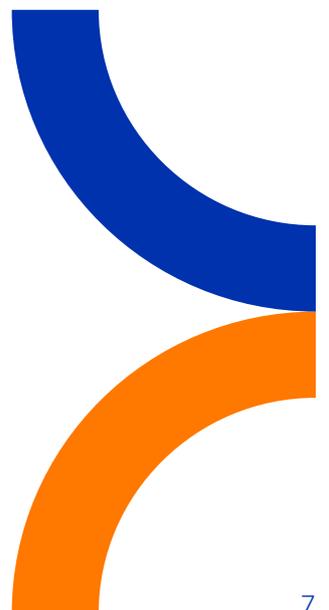

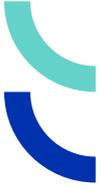

## Use Scenario 1: Bias on AI Algorithms

**Idea:**

- Explaining to students the bias that can be introduced in machine learning if data are not properly selected and analysed, and how it can have a relevant impact on automatic decision systems.
- *Why:* to introduce them to the impact of AI.
- *How:* by means of a developing and analysing simple decision systems.

**Target level:** upper primary school and secondary school (ages 8 to 18).

**Topic:** machine learning

**Possible resources:**

*Name:* **A Fresh Squeeze on Data**

- **Target level:** ages 8-10
- **Link:** https://freshsqueezekids.com
- **Description:** This lesson is intended to provide student awareness of bias and their relationship with data. As a demonstration, this lesson allows the teacher and the students to explore and experiment with data bias. The teacher will explain why data is important, how to collect data, what is bias and its relationship with data. The lesson will conclude by asking students to imagine practical, real-life implications of data bias.

*Name:* **AI + Ethics Curriculum for Middle School**

- **Target level:** ages 10-14
- **Link:** https://docs.google.com/document/d/1e9wx9oBg7CR0s5O7YnYHVmX7H7pnITfoDxNdrSGkp60/
- **Description:** On page 29 of this pdf file, there is a detailed activity with the required materials. Students will have to use an online programming tool (teachable machine), which does not require any previous experience on programming.

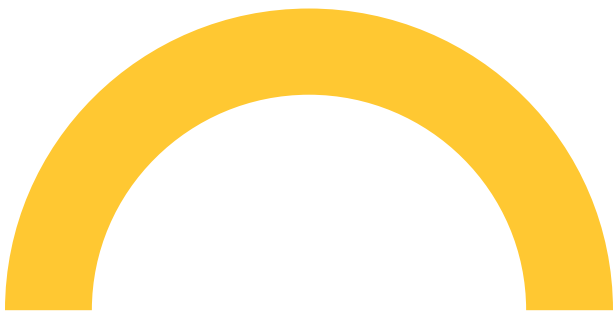
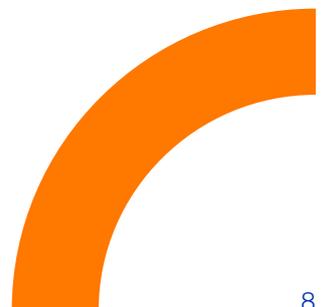

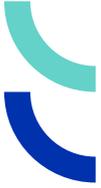

*Name:* **Our AI Code of Ethics**

- **Target level:** ages 10-14
- **Link:** https://studio.code.org/s/ai-ethics-2021/lessons/1
- **Description:** This lesson belongs to the code.org initiative, and it builds on prior activities involving research and finding sources to justify an argument. In small groups, students conduct research using articles and videos that expose ethical pitfalls in an AI area of their choice. Afterward, each group develops at least one solution-oriented principle that addresses their chosen area. These principles are then assembled into a class-wide "Our AI Code of Ethics" resource (e.g., a slide presentation, document, or webpage) for AI creators and legislators everywhere.

*Name:* **Build a Bot**

- **Target level:** ages 14-18
- **Link:** https://dschool.stanford.edu/resources/build-your-own-bot
- **Description:** This facilitation guide includes a set of activities for children, families, and parents to experiment with the potential and peril of AI assistants. In this document there are three workshops with facilitator guides, slide decks, worksheets, and other materials. These have all been designed as unplugged activities and do not require a computer.

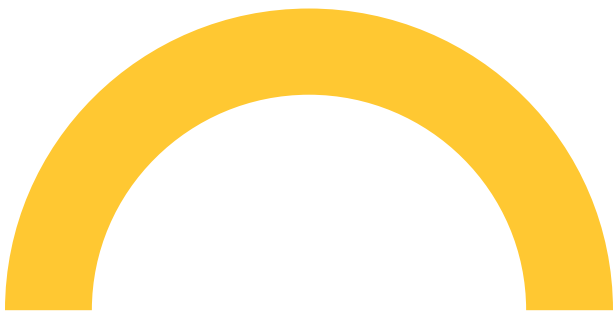
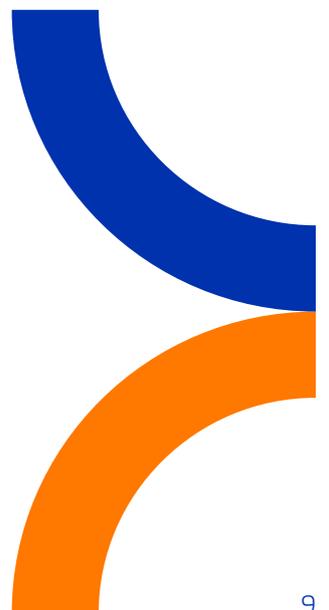

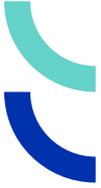

## Use Scenario 2: Computational Thinking and Algorithms

**Idea:**

- What is an algorithm? How does it work? How is it related to AI?
- *Why:* to introduce students in the fundamentals of problem solving by means of computer programs.
- *How:* by creating simple algorithms with and without AI.

**Target level:** lower secondary school (ages 10 to 14)

**Topic:** design thinking, problem solving, algorithms

**Possible resources:**

*Name:* **Algorithmic Literacy**

- **Link:** https://algorithmliteracy.org
- **Description:** Digital2030 (an experience by Digital Moment), the Canadian Commission for UNESCO (CCUNESCO) and UNESCO have partnered up to launch the Algorithm Literacy & Data Project to raise awareness and educate children about the presence of algorithms and how they influence our digital experiences — in other words, get algorithm literate. The goal is to empower children to exercise critical thinking in how they engage online, and to become proactive, creative users and makers rather than passive consumers.

*Name:* **AI + Ethics Curriculum for Middle School**

- **Link:** https://docs.google.com/document/d/1e9wx9oBg7CR0s5O7YnYHVmX7H7pnITfoDxNdrSGkp60/
- **Description:** On page 16 of this pdf file, there is a detailed activity with the required materials. Students will have to create an algorithm to make the "best" peanut butter and jelly sandwich. Students then explore what it means to be "best" and see how their opinions are reflected in their algorithms.

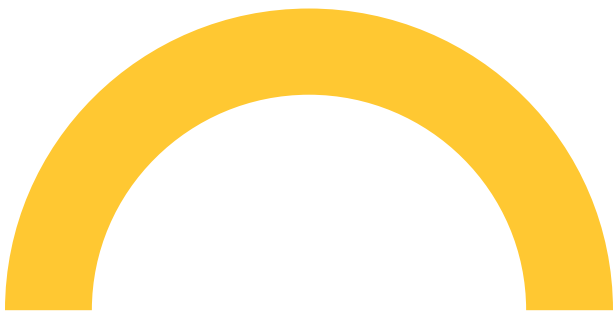
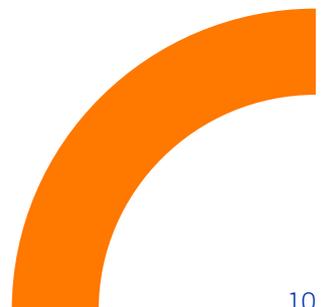

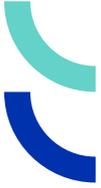

## Use Scenario 3: Data Privacy and AI

**Idea:**

- What data are AI systems collecting about us? How can we keep our online data private?
- *Why:* because AI changes the way some of our most basic human rights need to be protected and adapted.
- *How:* getting students to experiment with AI systems that collect personal data and reflect on what privacy means in the context of AI.

**Target level:** lower secondary school (ages 10 to 14)

**Topic:** privacy, digital rights, algorithms, bias

**Possible resources:**

*Name:* **Artificial Intelligence and Human Rights – Lesson 4 – Technology and Privacy**

- **Link:** https://www.dayofai.org/curriculum
- **Description:** AI Blueprint Bill of Rights (please register to use lessons for free). Students learn about the increasing use of AI in our everyday lives, and how that use is forcing us to consider how some of our most basic human rights need to be protected and adapted. Students focus on four specific rights: non-discrimination, privacy, transparency and safety, with a targeted lesson on each.

*Name:* **How Normal am I?**

- **Link:** https://www.hownormalami.eu/
- **Description:** Interactive documentary about how websites and cameras can collect personal and behavioural data and infer pieces of information such as your age, life expectancy, beauty score, body mass index, concentration level, and others. (Sherpa Horizon 2020 Project).

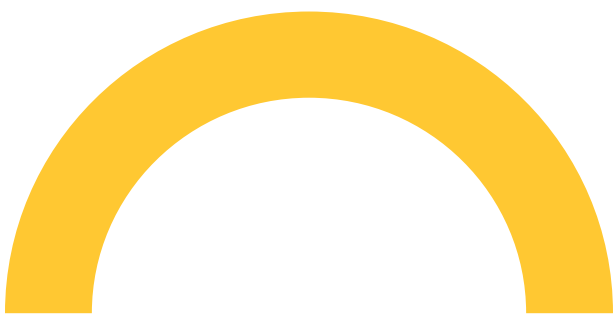
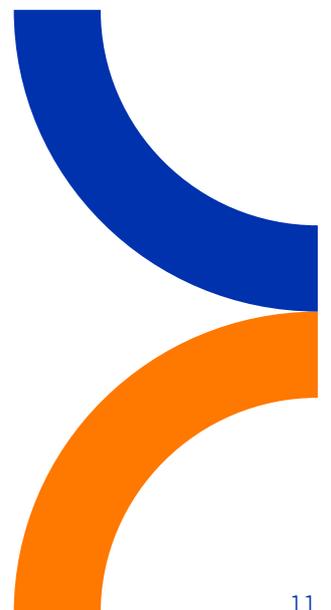

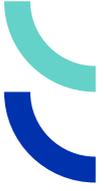

# Teaching *about* AI

In this section we describe projects and learning scenarios for primary and secondary schools (K-12) that provide examples for teaching about AI, to train students in the fundamentals of AI from an AI developer perspective. In most cases, these resources imply that students program a simple AI-based solution, with the aim of gaining the AI basics from the background. It is a more technical approach to AI education. Teaching about AI should cover an introduction to AI, which includes the definition and history as well as the importance and applications of AI in various fields. Four key areas of AI should be covered: perception and actuation, representation and reasoning, learning and the impact of AI. Students should get a clear idea of an AI system as an agent (computational system) that is situated in an environment (real or virtual) which interacts with it in an autonomous fashion (no human supervision required) to reach some design goals. Consequently, the teacher should differentiate AI from machine learning, which is a specific area of AI.

However, it is essential to have an introduction to machine learning, covering supervised, unsupervised, and reinforcement learning, as common machine learning algorithms such as linear regression, decision trees, clustering, and neural networks. It is important to explain different ways of data collection and organisation, including identifying relevant data sources for AI projects, data cleaning and pre-processing techniques. It is also important to introduce students in AI applications like computer vision: object detection, recognition, and segmentation, face recognition, and facial landmarks. Additionally, an overview of robotics and control systems should be provided. The topics should also include generative models, natural language processing (NLP) and its applications. The teacher should cover the social implications and biases of AI, privacy, and data security, AI in decision-making and policy, responsible AI development. Within this scope, it is necessary that educators and learners have a proper background in mathematics, programming, statistics and informatics. Education in this realm requires background knowledge about mathematics, statistics, informatics and programming.

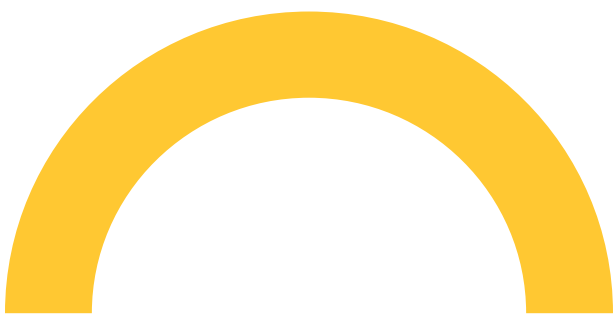
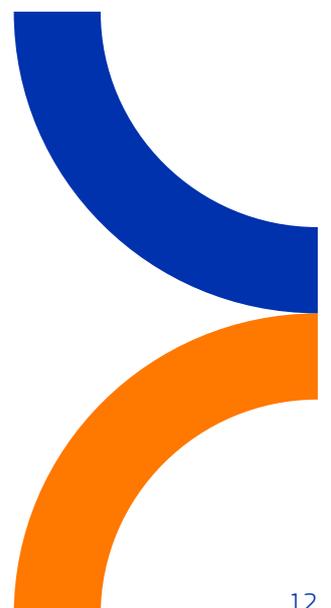

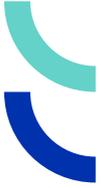

## Use Scenario 1: Representation & Reasoning in AI

**Idea:**

- Providing students with the fundamentals of representation and reasoning on AI by means of the development of a smartphone application using App Inventor.
- *Why:* to train them about these 2 key topics of AI.
- *How*: by means of developing a smartphone app

**Target level:** high school (ages 15 to 18)

**Topic:** representation and reasoning

**Possible resources:**

*Name:* **AI + Project**

- **Link:** https://drive.google.com/drive/u/1/folders/190-rLvuWvKsNtX7Mb2FOB0SLuZTx404w
- **Description:** The linked folder contains a guide file for teachers so they can implement this activity, as well as the code with the solution. Students will develop an app called "School Path Guide", using a graph representation and a simple probabilistic reasoning algorithm.

*Name:* **Elements of AI – Building AI – Dealing with Uncertainty**

- **Link:** https://buildingai.elementsofai.com/Dealing-with-Uncertainty
- **Description:** One of the reasons why modern AI methods actually work in the real world - as opposed to most of the earlier old-fashioned methods in the 1960-1980s - is the ability to deal with uncertainty. This activity shows students how Bayes probability works and how relevant it is for reasoning and problem solving in real AI.

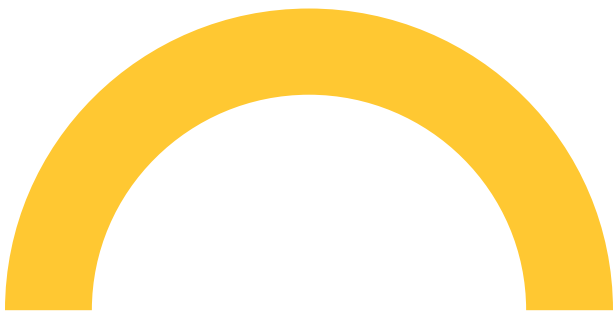
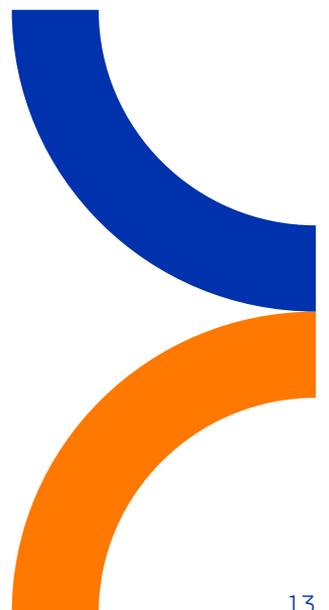

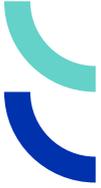

## Use Scenario 2: Recommendation Systems

**Idea:**

- Learning about "clustering" and using a clustering application to create their own recommendation system.
- *Why:* to understand how the use of clustering and filtering in recommendation systems in social media platforms can speed the spread of misinformation.
- *How:* by developing a basic understanding of what AI is and how it works in recommendation systems – what data it looks at, and how it chooses to filter content for us.

**Target level:** high school (ages 16 to 18)

**Topic:** recommendation, clustering, filter bubbles

**Possible resources:**

*Name:* **Day of AI – AI and Social Media**

- **Link:** https://www.dayofai.org/curriculum
- **Description:** AI in social media (please register to use lessons for free). Recommendation systems, the foundation of feeds and suggestions across social media platforms, define what we are and are not exposed to online. In this activity, students look at different forms of misinformation, how the use of clustering and filtering in recommendation systems in social media platforms can accelerate the spread of misinformation, and explore the social implications of these filter decisions for us as individuals, and as a society.

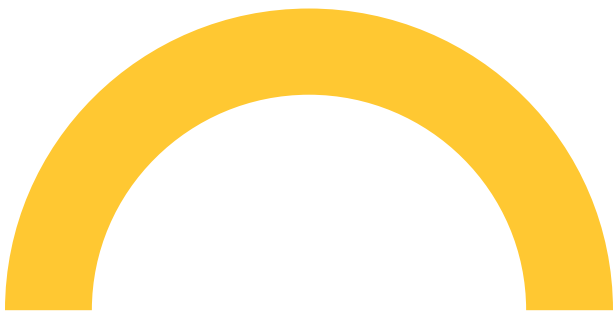
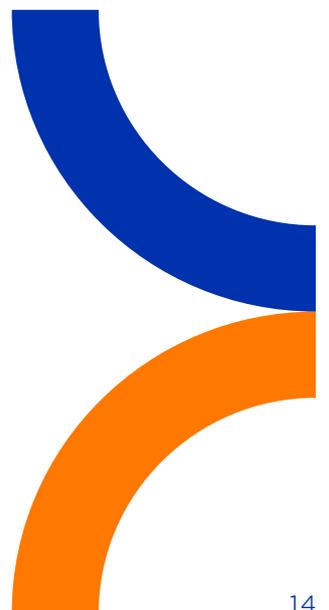

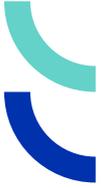

## Use Scenario 3: Teaching Machines to Classify

**Idea:**

- What is a classification task in the context of supervised machine learning context and what are the basic building blocks of a classification pipeline?
- *Why:* to recognise the importance of the quality and the quantity of training data and their impact on the accuracy and fairness of classifiers.
- *How:* by the implementation of a three-step procedure to build a classifier, test it, assess its performance in terms of accuracy and fairness and streamline it when possible.

**Target level:** lower secondary school and high school (ages 12 to 18)

**Topic:** classification, machine learning, datasets

**Possible resources:**

*Name:* **EU CodeWeek – Supervised Machine Learning**

- **Link:** https://codeweek.eu/training/introduction-to-artificial-intelligence-in-the-classroom and Teaching Machines to Classify: Intro to Supervised Machine Learning, for Lower Secondary School
- **Description:** This activity covers essential concepts of machine learning at an introductory level, focusing specifically on the task of classification. It aims to inspire the future generation of innovators to harness the potential of machine learning and AI and understand related advantages and limitations, through simple yet powerful case studies.

*Name:* **AI + Project – Image Classification with Machine Learning**

- **Link:** https://drive.google.com/drive/u/1/folders/1lohSLWg8yRsZQEiM2X2oW6Ne_hDYqI6M
- **Description:** Develop a smartphone app using App Inventor that allows to play a scavenger hunt searching game in school based on machine learning.

Scenarios have been developed with pre-university levels in mind, to be used by teachers in formal or informal education. If we move to higher levels, each speciality would require a different focus of teaching about AI. In general, specific training in AI is recommended. For general citizenship education (adult education, informal education), teaching about and for AI is encouraged, and some remarkable initiatives are already available like ciutadanIA or Elements of AI.

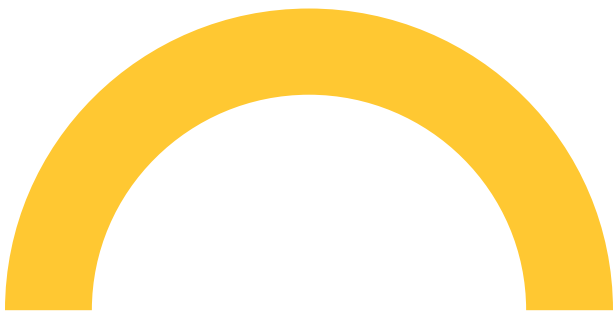
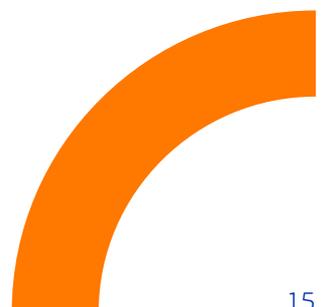

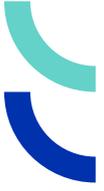

# Teaching *with* AI

In this section we describe use scenarios for learning/teaching with the use of AI-based tools. To appropriately apply AI-based tools in the classroom, it is important for the teacher to possess a basic knowledge of AI's capabilities, constraints and ethical considerations. As with any digital tool, it is essential to know how to select tools that align with the curriculum, pedagogical goals, and students' requirements, while considering the efficacy, ease of use, and privacy issues associated with these tools. To support and enrich the learning experience, educators should incorporate AI tools into their lesson plans while considering how AI can be used for teaching and for supporting students' learning and assessment, e.g., to personalise learning, provide feedback, or improve collaboration between peers. It is important for teachers to be aware of ethical implications such as bias, fairness, accountability, explicability and transparency of AI, also to assess regularly the effectiveness of AI tools used in the classroom and to ensure that learning objectives are being met and educational experiences are being enhanced. Collaboration between teachers, promoting the sharing of experiences, ideas, and best practices for using AI in the classroom can lead to more effective implementation of AI and other digital tools, as well as a better understanding of its potential benefits and limitations.

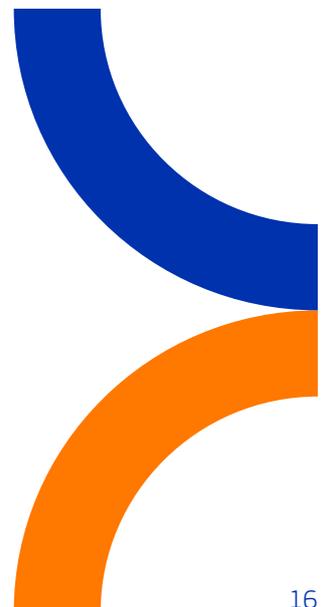

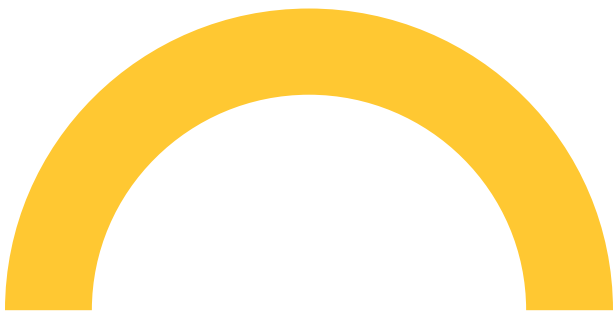

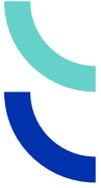

## Use Scenario 1: Automatic Content Generation

**Idea:**

- Taking advantage of generative AI tools
- *Why:* to create initial versions of documents, images or art that can be later improved by students.
- *How:* by means of generative AI tools

**Target level:** all

**Topic:** text creation, language, arts, maths

**Possible resources:**

*Name:* **Genial.ly Generative AI in Education**

- **Link:** https://view.genial.ly/63ec8abdc804dc0018561bbe?fbclid=IwAR3FfHFK_hWIsiVoW8GFjYaLeL8XZfkKfILVB94oiKumKpUohLL0AG_lxgk
- **Description:** Collection of content creation tools.

*Name:* **Language Styles**

- **Link:** https://chat.openai.com and https://you.com
- **Description:** Write a first draft using you.com or ChatGPT. Use the chosen AI to improve content. Then compare with the other groups, what must be changed to switch between different writing styles. Similar approach for foreign languages: to enrich vocabulary, find synonyms, etc, use Google/Microsoft Speech-to-Text for pronouncing exercises of new vocabulary.

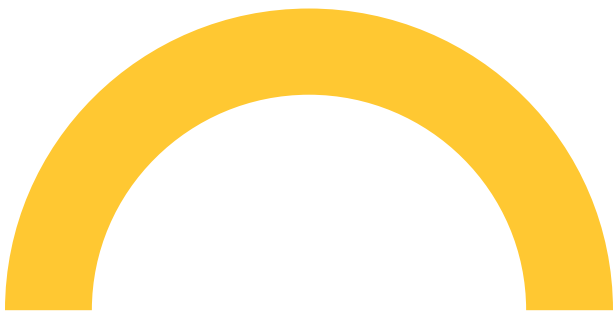
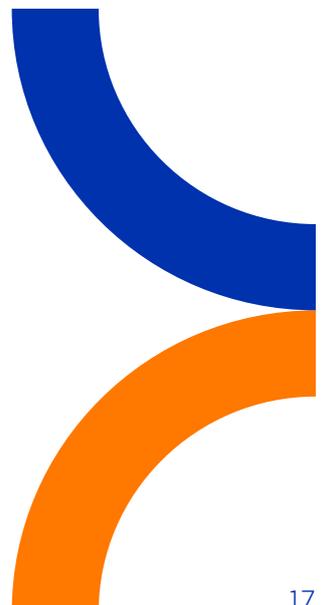

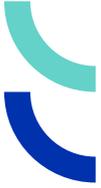

*Name:* **Text to Images**
- **Link:** https://labs.openai.com and https://you.com/imagine
- **Description:** For science: classification of animals or fruits. Exploring characteristics of animals or fruits more in depth and use that knowledge to train teachable machine to classify them.
  For language: create a scene from a story. Be aware of the descriptions given.
  For art: Use https://digitalcurator.art/ to create a gallery of a certain style or period. Explore the characteristics of the paintings. Use DALL-E to create your own in the chosen style, confront in group, use a teachable machine to train.

*Name:* **Teachable Machine**
- **Link:** https://teachablemachine.withgoogle.com
- **Description:** Simple visual tool to create machine learning models
  Biology, categorisation, of birds, insects, etc. and text to image
  Introduction of dynamism (e.g., Giacomo Balla - "Dynamism of a Dog on a Leash", 1912) as well as physical education and anatomy.

*Name:* **Learning with Quizlet**
- **Link:** https://quizlet.com
- **Description:** Doing flashcards, quizzed with Quizlet or other apps. Creation of flashcards and using them for learning is important.

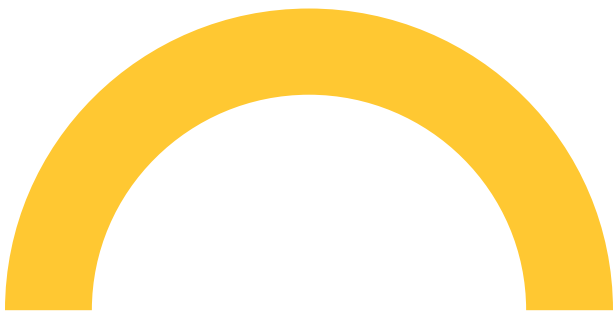
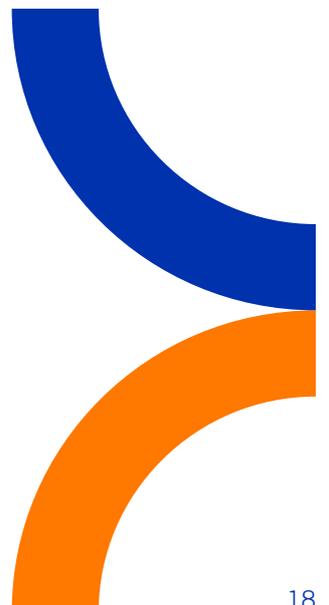

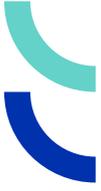

## Use Scenario 2: Intelligent Tutoring Systems

**Idea:**

- Support teachers on equalising the advance level of students in different subjects.
- *Why:* to simplify the management of heterogeneous groups.
- *How:* by using a learning platform that provides individualised monitoring

**Target level:** secondary school

**Topic:** mathematics, programming

**Possible resources:**

*Name:* **GOORU NAVIGATOR**

- **Link:** https://gooru.org/about/navigator/
- **Description:** Navigator is an intelligent tutoring system, which monitors students' training in different subjects to reach desired learning outcomes. It provides personalised materials to students.
- **Demo video:** https://www.youtube.com/watch?v=c1IXDN952GM
- **Demo video for K12 maths:** https://youtu.be/IkaAjce1l28
- **NOTE:** It is not free (https://gooru.org/about/pricing/).

*Name:* **DOMOSCIO SPARK**

- **Link:** https://domoscio.com/en/domoscio-spark-2/
- **Description:** An intelligent tutoring system, which monitors students' training in different subjects to reach desired learning outcomes. It provides personalised materials to students.
- **Demo video:** https://www.youtube.com/watch?v=3LygEeV-NhQ
- **NOTE:** It is not free.

*Name:* **COBIE AI**

- **Link:** https://cobie.io/smart-classroom/
- **Description:** Intelligent tutoring system that helps students learn to code. The teacher monitors what students are doing and provides help if needed to multiple students at the same time with the help of Cobie AI assistant. The system also includes a lecture synthesis system so that teachers can easily and quickly create personalised lectures.

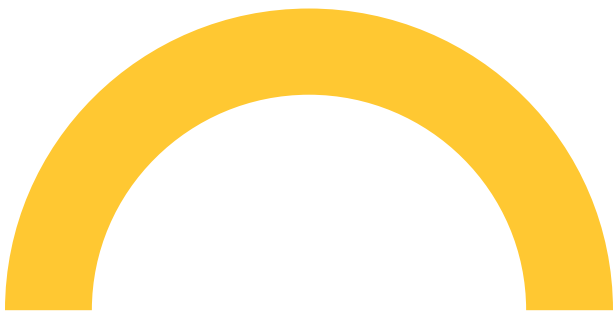
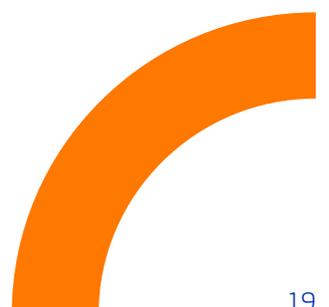

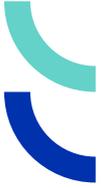

## Use Scenario 3: Automatic Translation of Conversations

**Idea:**

- Showing students how to maintain a conversation in real time in two languages, to communicate with other persons quickly using the smartphone.
- *Why*: to show them that language knowledge should not be an obstacle for socialising.
- *How*: by means of AI-based technology and their own smartphone.

**Target level:** all

**Topic:** conversation, language

**Possible resources:**

*Name:* **SAYHI**

- **Link:** https://www.sayhi.com/en/translate
- **Description:** It is a smartphone app that detects your speech in any language and translates it to any other language in real time.

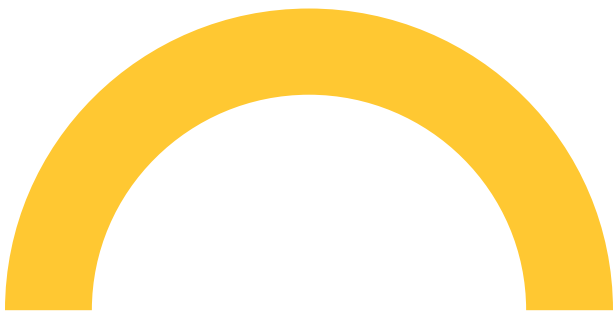
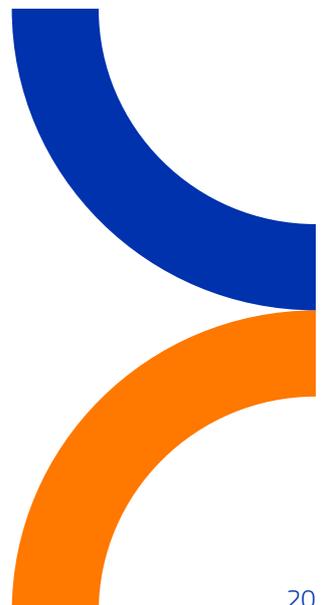

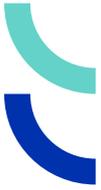

# Recommendations by the Squad

The following diagram summarises the conceptual organisation of AIEd teaching categories presented in this report:

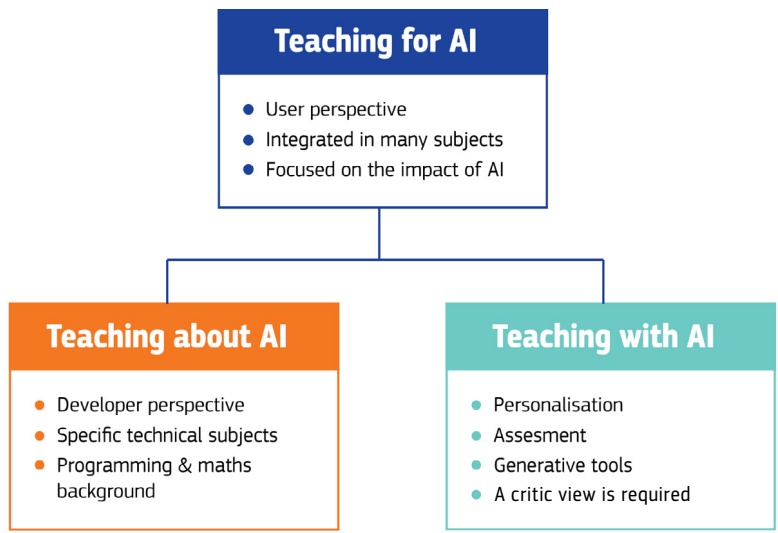

It can be observed that our main recommendation places Teaching *for* AI is on the top, representing the idea that it should be a pre-requisite for the other two.

Focus first on teaching *for* AI by means of practical projects and learning scenarios that provide activities for teachers to engage students in activities that improve knowledge, skills, and attitudes towards how AI systems are used in today's society and focus on everyday application that are driven by AI.

Take advantage of existing resources for teaching *with* AI to enhance teaching and learning. It is essential to know how to select tools that align with the curriculum, pedagogical goals, and students' requirements, while considering the efficacy, ease of use, and privacy issues associated with these tools.

Apply a developer approach when teaching *about* AI to train more specialised students in the fundamental areas of real-world AI, like perception, reasoning, representation or learning. They must face different AI challenges through hands-on and programming projects, so they attain the AI basics from a more technical perspective.

*Members of the EDEH squad on artificial intelligence in education who dedicated time for this briefing report: Dara Cassidy, Yann-Aël Le Borgne, Francisco Bellas, Riina Vuorikari, Elise Rondin, Madhumalti Sharma, Jessica Niewint-Gori, Johanna Gröpler, Anne Gilleran and Lidija Kralj.*

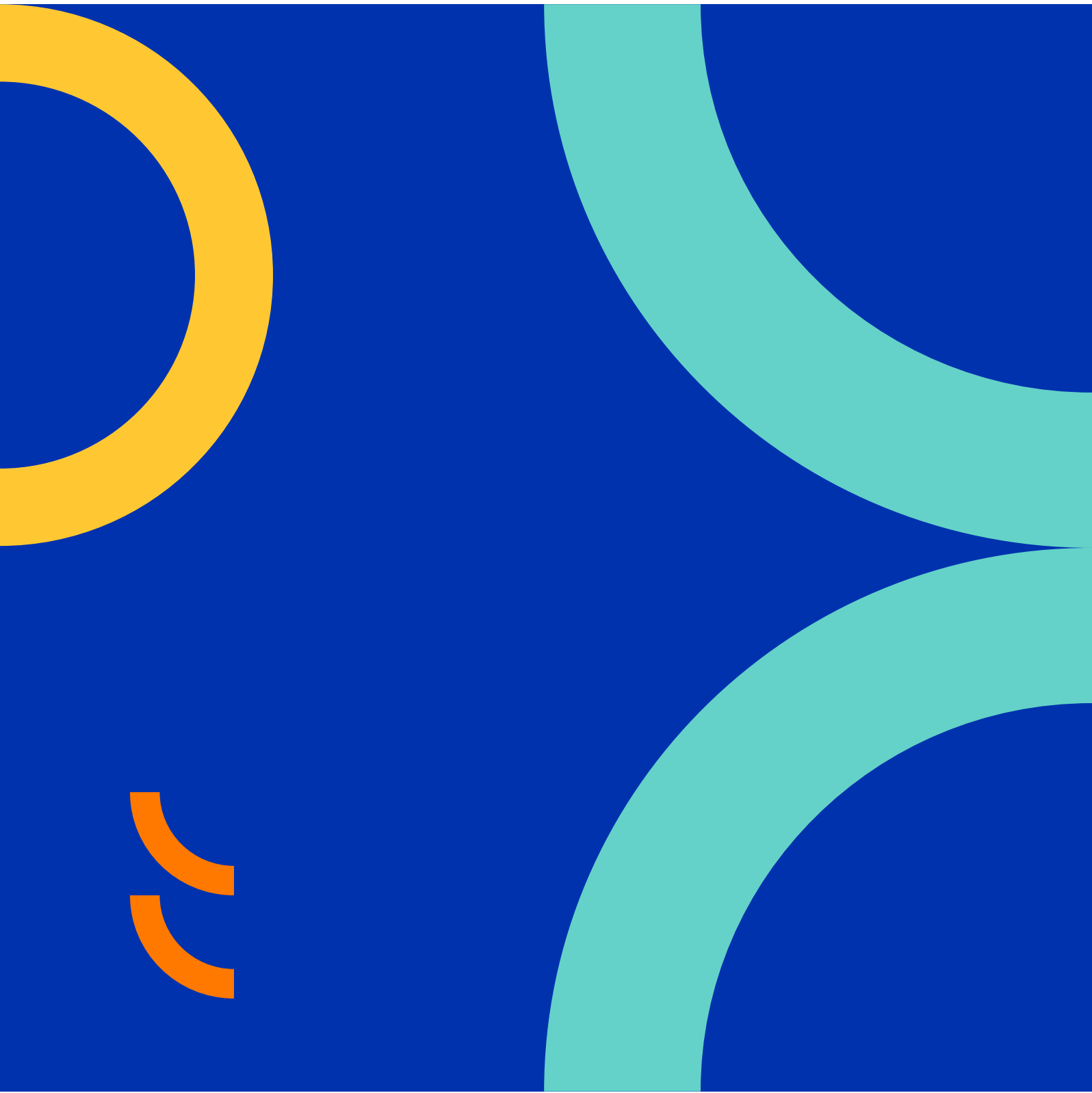